\documentclass[showpacs,preprintnumbers,amsmath,amssymb,superscriptaddress,nofootinbib,english]{revtex4}

\usepackage{epstopdf}
\usepackage{graphicx}
\usepackage{dcolumn}
\usepackage{bm}
\usepackage{epsfig}
\usepackage{graphicx}
\usepackage{hyperref}
\usepackage[usenames]{color}
\usepackage{url}

\hypersetup{
    colorlinks=true,
    linkcolor=green,
    citecolor=blue,
}

\newcommand{\remove}[1]{}

\def\be{\begin{equation}}
\def\ee{\end{equation}}
\def\ba{\begin{eqnarray}}
\def\ea{\end{eqnarray}}

\begin{document}

\title{Reheating and preheating in the simplest extension of Starobinsky inflation}
\author{Carsten van de Bruck}
\email[Email address: ]{C.vandeBruck@sheffield.ac.uk}
\affiliation{Consortium for Fundamental Physics, School of Mathematics and Statistics, University of Sheffield, Hounsfield Road, Sheffield, S3 7RH, United Kingdom}

\author{Peter Dunsby}
\email{peter.dunsby@uct.ac.za}
\affiliation{Cosmology and Gravity Centre (ACGC), Department of Mathematics and Applied Mathematics, University of Cape Town, Rondebosch 7701, South Africa}

\author{Laura E. Paduraru}
\email[Email address: ]{LEPaduraru1@sheffield.ac.uk}
\affiliation{Consortium for Fundamental Physics, School of Mathematics and Statistics, University of Sheffield, Hounsfield Road, Sheffield, S3 7RH, United Kingdom}

\date{\today}

\begin{abstract}
The epochs of reheating and preheating are studied in a simple extension of the Starobinsky inflationary model, which consists of an $R^2$--correction to the Einstein--Hilbert action and an additional scalar field. We find that if the $R^2$--correction at the end of inflation is dynamically important, it affects the expansion rate and as a consequence the reheating and preheating processes. While we find that the reheating temperature and duration of reheating are only slightly affected, the effect has to be taken into account when comparing the theory to data. In the case of preheating, the gravitational corrections can significantly affect the decay of the second field. Particle production is strongly affected for certain values of the parameters in the theory.
 \end{abstract}

\maketitle
\section{Introduction}
Inflation was introduced as an extension to the standard Hot Big Bang cosmological model, based on a Friedmann-Robertson-Walker (FRW) metric, in order to resolve a number of problems which arose when trying to reconcile this model with observations of the universe \cite{Guth:1980zm,Linde:1981mu,Albrecht:1982wi}. Inflation is a phase of accelerated expansion in the very early history of the Universe, which allows for near flat spatial curvature and homogeneity in our epoch and for quantum fluctuations to be generated which became the seeds for the observed structures in the universe \cite{Guth:1982ec}, \cite{Starobinsky:1982ee}. In most models this phase is driven by one (see \cite{Kolb:1999ar} for a classification of single-field models) or several scalar fields, whose energy dominates the energy-momentum tensor. Since the inflaton (the scalar field responsible for driving inflation) has not yet been discovered, the inflationary paradigm can be viewed as a purely theoretical fix, however there are working models of this framework which can be embedded into fundamental theories. In other models, such as the one introduced by Starobinsky, inflation is caused by an $R^2$--modification of the Einstein--Hilbert action of General Relativity \cite{Starobinsky:1980te}. Examples of how these can be embedded in fundamental theories are found in \cite{Sebastiani:2013eqa}, \cite{Amoros:2014tha}.

Reheating is the process through which particles are created at the end of the inflationary phase (\cite{Traschen:1990sw}, \cite{Dolgov:1990}), which occurs by coupling the inflaton fields to matter. Such couplings arise via the gravitational sector or the matter sector \cite{Bassett:2005xm}, \cite{Allahverdi:2010xz}. Soon after the end of inflation, the inflaton fields begin to oscillate around the minimum of their effective potentials, producing particles, which interact with each other and reach thermal equilibrium at $T_r$, the reheating temperature. 

In the elementary theory of reheating, the oscillating inflaton fields produces radiation via the tree--level decay of inflaton particles into relativistic particles. Reheating completes around the time when the rate of expansion of the universe becomes smaller than the total decay rate of the inflaton into new fields. It was pointed out, however, that perturbative reheating is not the full story. Instead, particle production can occur via parametric resonance (see \cite{Kofman:1994rk,Shtanov:1994ce,Yoshimura:1995gc,Khlebnikov:1996mc,Kofman:1997yn}). This process is called preheating and is non--perturbative in nature. Whether or not preheating will occur depends on the details of the couplings between the fields in the theory, but many examples have been found in which preheating is very effective. 

In the following analysis, we consider one specific model for inflation, which is an extension of the model first introduced by Starobinsky \cite{Starobinsky:1980te}. Starobinsky inflation is based on a modification to the Einstein-Hilbert action, which involves adding the square of the Ricci scalar in the action and it belongs to the larger class of $f(R)$ theories of gravity. In the original work \cite{Starobinsky:1980te}, the $R^2$--correction came from quantum corrections to the Einstein--Hilbert action. We extend the original Starobinsky model by adding a scalar field, which we call $\chi$ \cite{Gottlober:1990um,vandeBruck:2015xpa}. The quantum correction to gravity is dynamically equivalent to a scalar field we call $\phi$, which drives inflation together with the scalar $\chi$. We find that at the end of inflation, the $R^2$--correction can still affect the evolution of the Hubble expansion rate $H$, thus interfering in the reheating process, even if its decay is subdominant compared to the decay of $\chi$. In this paper we will investigate this issue in detail.

In \cite{vandeBruck:2015xpa}, the inflationary dynamics of the above model was studied in the Einstein frame and compared to the Planck 2015 data. However, for the analysis of the post-inflationary epoch presented in the present paper, we chose to work in the Jordan frame. This is motivated by the standard form the coupling terms take in this case, which makes  the physics of the system easier to interpret. The paper is organised as follows: In Section \ref{sec:theory} we define our model and write down the equations of motion in the Jordan frame. In Section \ref{sec:reheating} we study the process of reheating. Preheating is discussed in Section \ref{sec:preheating}. Finally, our conclusions can be found in Section \ref{sec:conclusions}. 

Unless otherwise specified, natural units will be used throughout this paper and Greek indices run from 0 to 3. The symbol $\nabla$ represents the usual covariant derivative and $\partial$ corresponds to partial differentiation and $\Box = \nabla^\mu\nabla_\mu$. We use the $-,+,+,+$ signature and the Riemann tensor and Ricci tensor are defined in the standard way:
\begin{eqnarray}
R^{\alpha}{}_{\beta\mu\nu}=\Gamma^{\alpha}{}_{\beta\nu,\mu}-\Gamma^{\alpha}{}_{\beta\mu,\nu}+ \Gamma^{\delta}{}_{\beta\nu}\Gamma^{\alpha}{}_{\mu\delta}-\Gamma^{\delta}{}_{\beta\mu}\Gamma^{\alpha}{}_{\nu\delta}\;,~R_{\alpha\beta}=g^{\mu\nu}R_{\alpha\mu\beta\nu}\;,
\end{eqnarray}
where $\Gamma^{\alpha}{}_{\beta\mu}$ are the usual Christoffel symbols.
\section{The model and inflationary dynamics}\label{sec:theory}
The analysis of re-- and preheating in our extension of the Starobinsky model will be performed in the Jordan frame, so that the decay rates can be calculated and defined in a standard way.  In this frame the model is specified by the action 
\begin{equation}
\begin{split}
S &= \frac{1}{2\kappa}\int d^{4}x \sqrt{-g} \bigg[ R+ \mu R^2 \bigg] + \int d^{4}x \sqrt{-g}\bigg[ - \frac{1}{2} g^{\mu \nu}\partial_{\mu}\chi \partial_{\nu}\chi 
- \frac{1}{2} m_\chi^2 \chi^{2} \bigg] + {\cal S}_{\rm Other}~.
\end{split}
\end{equation}
Here $\kappa = M_{\rm Pl}^{-2}$, where $ M_{\rm Pl}$ is the reduced Planck mass and the parameter $\mu$ has units [mass]$^{-2}$. We define ${\cal S}_{\rm Other}$ to be the part of the action for other forms of energy density in the universe; in the case below we assume this corresponds to relativistic particles with an equation of state $p/\rho=1/3$ (radiation) produced by the decay of $\chi$ after inflation. The equations of motion for a $f(R)$ theory have been derived many times in the literature (see e.g., \cite{Guo:2013swa}), they read 
\begin{eqnarray}
f' R_{\mu\nu} - \frac{1}{2} f g_{\mu\nu} - \left( \nabla_\mu \nabla_\nu - g_{\mu\nu} \Box \right) f' = \kappa T_{\mu\nu}\;,
\end{eqnarray}
with $f' = df/dR$. The energy momentum tensor $T_{\mu\nu}$ includes the contribution of $\chi$. The trace of this equation results in an equation for $f'$, given by  
\begin{eqnarray}
\Box f' = \frac{1}{3} \left(2f - f' R\right) + \frac{\kappa}{3} T\;,
\end{eqnarray}
where $T$ is the trace of $T_{\mu\nu}$. This motivates the introduction of the scalar $\phi = f'$ with a potential defined as  
\begin{equation}\label{pot}
\frac{dV}{d\phi} = \frac{1}{3}\left( 2f - \phi R \right)\;,
\end{equation}
so that the equation for $\phi$ is given by 
\begin{equation}
\Box \phi = V' + \frac{\kappa}{3} T\;.
\end{equation}
We define the mass of the scalar degree of freedom associated with the $f(R)$--correction to General Relativity to be $ m_\phi^2 \equiv V'' = d^2 V/d\phi^2 = 1/6\mu$. 
Specialising to the case of the FRW metric, $ds^2 = -dt^2 +a^2(t)d{\bf x}^2$, and with a given choice of $f(R) = R + \mu R^2$, the field equations read, with $H = \dot a/ a$,  
\begin{eqnarray}
\ddot \phi &+& 3H \dot\phi + V' = \frac{\kappa}{3}\left(\rho_\chi - 3p_\chi \right), \label{eq:phi} \\
\dot H &=& \frac{R}{6} - 2 H^2 = \frac{\phi - 1}{12\mu} - 2H^2 \label{eq:dotH} \\ 
H^2 &=& \frac{\kappa}{3\phi} \rho - \frac{f-\phi R}{6\phi} - \frac{\dot\phi}{\phi}H, \\
\ddot \chi &+& 3H\dot\chi + m_\chi^2 \chi = - \Gamma \dot\chi, \\
\dot\rho_r &=& -4H \rho_r + \Gamma \dot \chi^2\;.\label{eq:rhodot}
\end{eqnarray}
In these questions, $\rho_\chi$ and $p_\chi$ are the energy density and pressure of the $\chi$--field, respectively, $\rho_r$ is the density of the radiation produced due to the decay of $\chi$ and $\Gamma$ is the decay rate of $\chi$ into radiation. 
We will focus our analysis on the cases where both fields oscillate during reheating. This implies that the mass ratio $m_\chi / m_\phi$ does not differ much from one. Assuming that the decay of $\chi$ dominates the decay of $\phi$, we do not introduce a decay term for the field $\phi$\footnote{After quantising the gravitational sector, the field $\phi$ is allowed to decay into other particles, but the decay rate is suppressed by powers of $M_{\rm Pl}$. For the mass range for $\phi$ and $\chi$ we are considering in this paper, our assumption that the decay of $\chi$ dominates that of $\phi$ is justified.}. Thus, for the purpose of this paper, the role of $\phi$ is to affect the evolution of the Hubble parameter $H$, which in turn will affect the behaviour of $\chi$. 

\section{Reheating}\label{sec:reheating}

In this section we study the process of (perturbative) reheating in the model.  As already mentioned, our assumption is that the field $\chi$ dominates the production of relativistic particles, whereas any decay of the field $\phi$ is negligible. Consequently, the decay terms appear only in the equations for $\rho_r$ and $\chi$,  see eq (\ref{eq:phi})--(\ref{eq:rhodot}). In light of our assumptions, both $\phi$ as well as $\chi$ oscillate around the minimum of their potential at the end of inflation. We therefore study the effect of the $R^2$--corrections on reheating, knowing that they do influence the predictions for inflation. These equations will be integrated from the beginning of inflation, which we take to last longer than 50 e-folds, until the end of reheating, at which point $\rho_r a^4 =$constant, i.e., the $\chi$-field has completely decayed into radiation. The reheating temperature is defined by 
\begin{equation}
\rho_r = \frac{\pi^2}{30}g_{\rm dof} T^4 \;,
\end{equation}
where $g_{\rm dof}\approx 100$ is the number of relativistic degrees of freedom. It is useful to define the e-fold averaged equation of state $w_{\rm Nav}$ as follows 
\begin{equation}
w_{\rm Nav} = \frac{1}{N_{er} - N_{ei}} \int_{N_{ei}}^{N_{er}} w dN\;,
\end{equation}
where the subscripts `$ei$' denotes the time at the end of inflation and `$er$' the time at the end of reheating; $w$ is the total equation of state, defined by 
\begin{equation}
-\frac{\dot H}{H^2} = \frac{3}{2}\left( 1 + w \right)\;.
\end{equation}
It is easy to show that \cite{vandeBruck:2016xvt}
\begin{eqnarray}
w_{\rm Nav} &=& \frac{2}{3} \frac{1}{\Delta N} \ln \left(\frac{H_{ei}}{H_{er}}  \right) - 1\;.
\end{eqnarray}
To be precise, we numerically define the end of reheating when $\rho_r/(\rho_\chi + \rho_r) > 0.9$. 
The numerical results are compiled in Table \ref{ta:reh} above. The choice of parameters are motivated by our previous work \cite{vandeBruck:2015xpa}, chosen such that the predictions in our model are in agreement with the Planck 2015 data \cite{Ade:2015xua}.

\begin{figure}[!htb]
 \minipage{0.32\textwidth}
  \includegraphics[width=\linewidth]{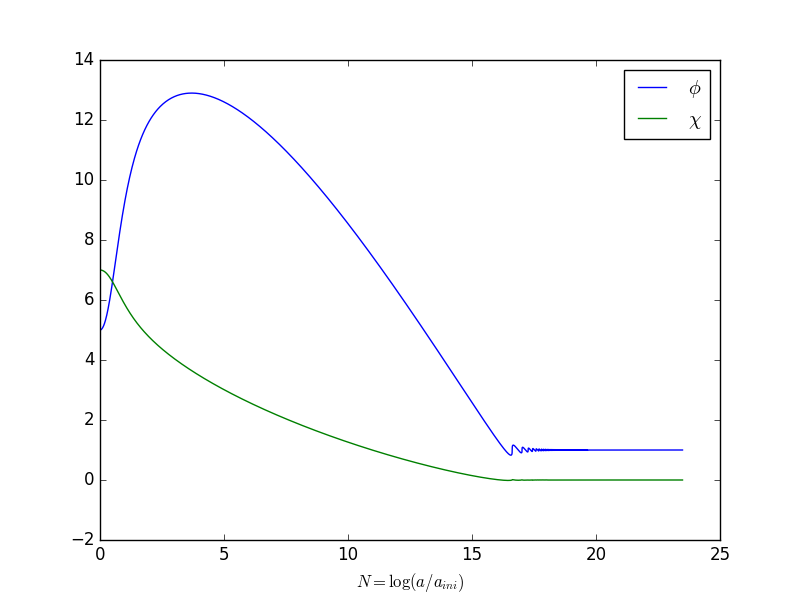}
  
\endminipage\hfill
\minipage{0.32\textwidth}
  \includegraphics[width=\linewidth]{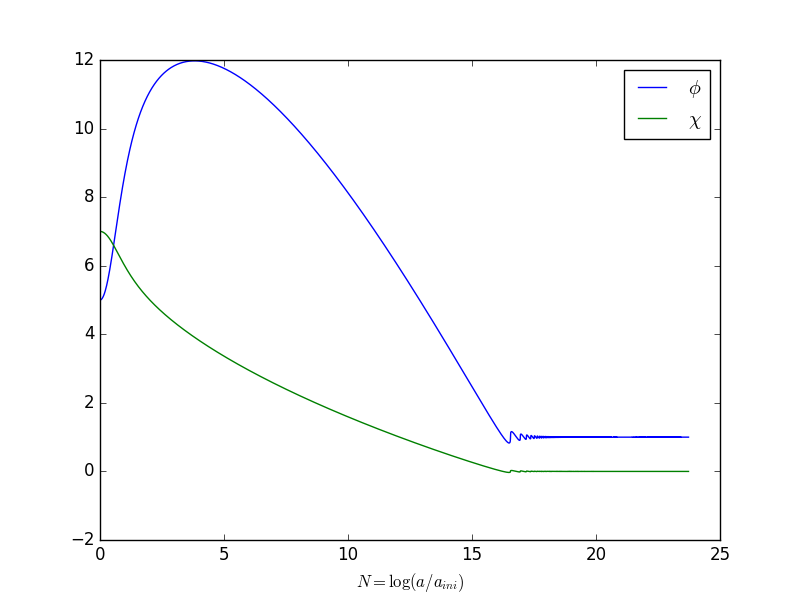}
  
\endminipage\hfill
\minipage{0.32\textwidth}%
  \includegraphics[width=\linewidth]{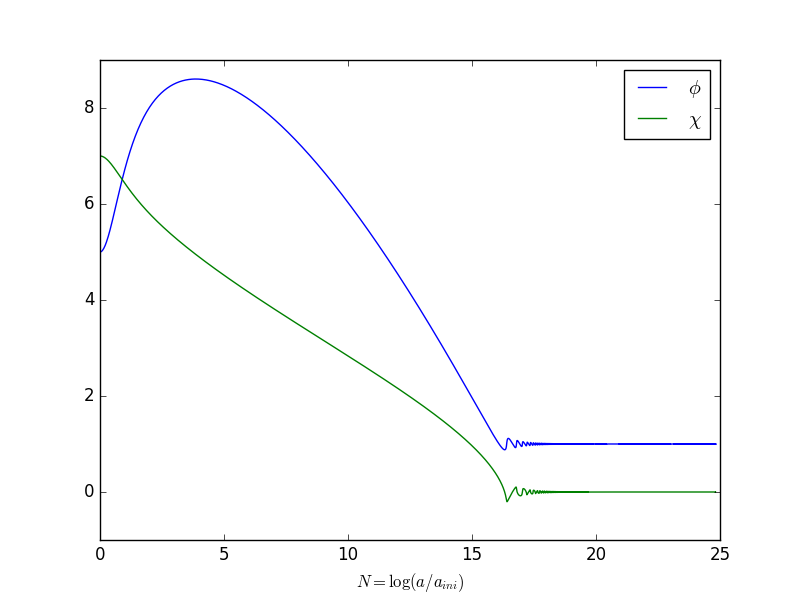}
  
  \endminipage
\caption{\label{fig:3runs} Inflationary trajectories for different mass ratios with identical initial conditions. The fields values are shown as a function of $e$--fold number $N=\log(a/a_{ini}$), where $a_{ini}$ is the value of the scale factor at the beginning of the run. On the left, we show the case for $m_\phi = 0.9m_\chi$, 
in the middle $m_\chi = m_\phi$ and on the right  $m_\phi = 1.5 m_\chi$. Here we have chosen $m_\chi = 1.3\times 10^{-6} M_{Pl}$ and $\Gamma = 10^{-3}m_\chi$. These figures are for illustrative purposes only to show the behaviour of the fields at the end of inflation.}

\end{figure}

In our following analysis we fix the value of the decay rate $\Gamma$ and  we consider combinations of masses of the $\chi$-field and values of the mass ratio which give predictions for the primordial power spectrum in agreement with the Plank 2015 data. As it can be seen, increasing $m_\phi$, which makes the $R^2$--corrections become less important, leads to a slight increase in the reheating temperature and the duration of reheating. On the other hand, decreasing $m_\phi$ and therefore increasing the importance of the $R^2$--correction, decreases the reheating temperature and shortens the reheating period. This can be understood physically by noting that, for $m_\phi<m_\chi$, $\chi$ will approach its minimum faster than $\phi$. As a result, any radiation produced by the decay of $\chi$ would be diluted away by the expansion of the universe as it is affected by the $\phi$ field.  We illustrate the evolution of the fields in Fig. \ref{fig:3runs}, where we choose identical initial conditions for three simulations with different mass ratios, i.e., $m_\phi<m_\chi$, $m_\phi=m_\chi$ and $m_\phi>m_\chi$.

\begin{table}

\begin{center}
    \begin{tabular}{| l | l | l | l | l |}
    \hline
    $m_\phi/m_\chi$ & $m_\chi (M_{Pl})$ &$T_{re} (GeV)$ & $w_{\rm Nav} $ & $\Delta N $\\ \hline
    $1.5$ & $5.89\cdot 10^{-6}$ & $2.11\cdot 10^{13}~$ & 0.0416 & 5.17 \\ \hline
    $1$ & $8.51\cdot 10^{-6}$ & $8.72\cdot 10^{12}~$ & 0.0066  & 4.95  \\ \hline
    $0.9$ & $9.33 \cdot 10^{-6}$ & $5.47\cdot 10^{12}~$ & 0.0998 & 4.86   \\ \hline
    \end{tabular}
\caption{\label{ta:reh} Reheating predictions for allowed values for $m_\phi$ and $m_\chi$. Here we set $\Gamma = 10^{-3}m_{\chi}$. $\Delta N = N_{er} - N_{ei}$ is the duration of the reheating phase. The end of reheating is defined to be the time when $\rho_r/(\rho_\chi + \rho_r) > 0.9$.}
\end{center}
\end{table}

In summary, we have found that the more important the $R^2$--corrections are at the end of inflation relative to the contribution from the $\chi$-field, the lower the reheating temperature and the shorter the reheating period. 

\section{Preheating}\label{sec:preheating}
\subsection{The view from the Jordan frame}

We turn now our attention to preheating in the extended Starobinsky model. We add an additional scalar field $\sigma$, which interacts directly with the $\chi$ field via a four-leg interaction term, so that the full action is given by 

\begin{equation}
\begin{split}
S &= \frac{1}{2\kappa}\int d^{4}x \sqrt{-g} \bigg[ R+ \mu R^2 \bigg] + \int d^{4}x \sqrt{-g}\bigg[ - \frac{1}{2} g^{\mu \nu}\partial_{\mu}\chi \partial_{\nu}\chi 
- \frac{1}{2} g^{\mu \nu}\partial_{\mu}\sigma \partial_{\nu}\sigma  - \frac{1}{2} m_\chi^2 \chi^{2} - \frac{1}{2} m_\sigma^2 \sigma^{2} - \frac{1}{2} h^2 \chi^2 \sigma^2 \bigg]\;.
\end{split}
\end{equation}

The $\sigma$ field is neglected during inflation, so we set its vacuum expectation value to be zero. The evolution of perturbations around the vacuum expectation value with momentum ${\bf k}$ obeys 
\begin{equation}
\ddot\sigma_{k} + 3 H \dot\sigma_k + \left( \frac{k^2}{a^2} + m_\sigma^2 + h^2 \chi^2 \right)\sigma_k = 0\;. 
\end{equation}
As it is well known, in the standard Einstein gravity case, for certain values of $k$, parametric resonance can occur \cite{Kofman:1994rk}, resulting in an explosive growth of the particle number density $n_k$, given by 
\begin{equation}\label{partnumber}
n_k = \frac{1}{2\omega_k}\left( |\dot\sigma_k |^2 + \omega_k^2 |\sigma_k|^2 \right) - \frac{1}{2}\;,
\end{equation}
where $\omega_k^2 = (k/a)^2 + m_\sigma^2 + h^2 \chi^2$.\footnote{The equation above is justified by $n_k = \rho_k /\omega_k$, where $\omega_k$ is the energy of the harmonic oscillator with mode $k$ and $\rho_k = \left( |\dot\sigma_k |^2 + \omega_k^2 |\sigma_k|^2 \right)/2 - \frac{1}{2}\omega_k$ is the energy density with subtracted ground state energy $\omega_k/2$.} We will now investigate whether this effect happens in the extended Starobinsky model. 

\begin{figure}[!htb]
 \minipage{0.5\textwidth}
  \includegraphics[width=\linewidth]{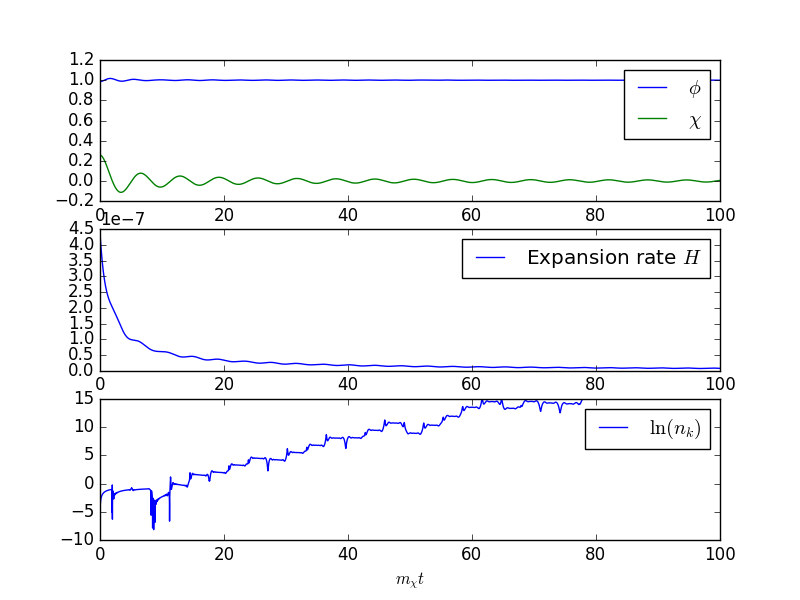}
\endminipage\hfill
\minipage{0.5\textwidth}
  \includegraphics[width=\linewidth]{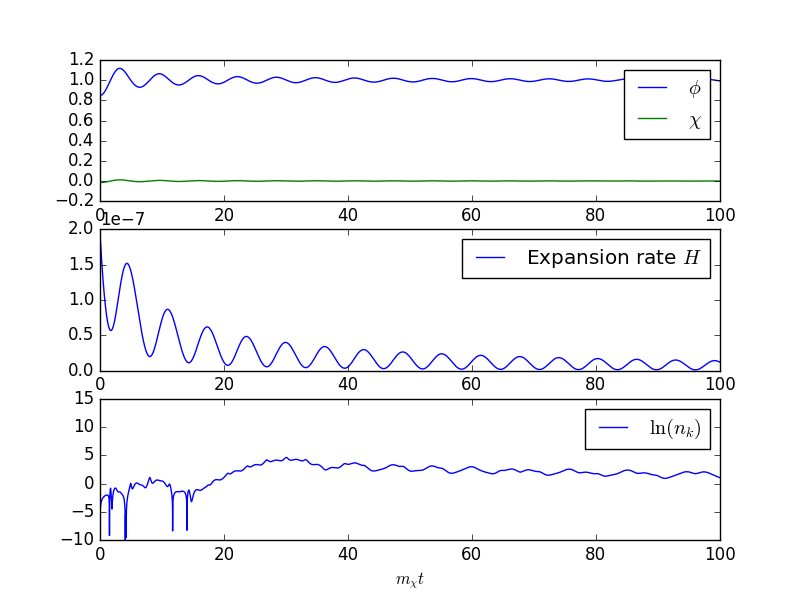}
\endminipage\hfill

\caption{\label{fig:preh_runs} Preheating for mass ratio $m_\phi / m_\chi = 1.5$ (left) and $m_\phi / m_\chi = 1.0$ (right). The upper panels show the evolution of $\phi$ and $\chi$ at the end of inflation, the middle panels show the evolution of the expansion rate $H$ and the lower panels show the particle number $n_k$, defined in eq. (\ref{partnumber}). As it can be seen, the mass ratio affects the evolution of $n_k$ significantly.}

\end{figure}

We numerically integrate the equations for two different cases with $m_\chi  = 1.3\cdot10^{-6}$M$_{\rm Pl}$,  $m_\sigma = 10^{-2}m_\chi$, $h =5\cdot10^{-4}$ and $k = 5\cdot 10^{-7}$.
In the first case,  we choose $m_\phi = 1.5m\chi$. The results are shown in the left panel of Fig. (\ref{fig:preh_runs}). Here, the $\phi$ field oscillates around $\phi_{\rm{min}} = 1~$M$_{\rm Pl}$, but the amplitude is rather small and therefore the modifications to General Relativity due to the $R^2$--corrections are not significant. The field $\chi$ oscillates around $0$, but with a much larger amplitude. As it can be seen, the particle number density $n_k$ of particles with momentum $k$, grows rapidly. Because $\phi\approx 1$, the dynamics of the field $\chi$ is very close to that of General Relativity. There are only minor deviations, due to the small oscillations of $\phi$ around 1~M$_{\rm Pl}$, affecting slightly the evolution of the expansion rate $H$. 

In the second case, we choose $m_\phi = m_\chi$. The results are shown in the right panel of  Fig. (\ref{fig:preh_runs}). In this case, the field $\phi$ oscillates around $\phi_{\rm{min}} = 1~$M$_{\rm Pl}$ with a much larger amplitude, whereas the $\chi$ field oscillates around 0 with a  smaller amplitude. As a consequence, the modifications to General Relativity are more important, with the expansion rate $H$ behaving  in an unconventional way and showing oscillatory behaviour, due to the oscillations of the $\phi$ field, see eq. (\ref{eq:dotH}). As a result, the number density $n_k$ in this second case does not exhibit much growth. 

In summary, we have found that preheating is much less efficient if the $R^2$--corrections are large at the end of inflation. We attribute this to the impact of these corrections to the evolution of the expansion rate $H$, which in turn affects the evolution of $\chi$.
 
\subsection{The view from the Einstein frame}

It is illuminating to consider the physics of this problem from the Einstein frame point of view.  Note that, in the Jordan frame, we cannot neglect the expansion of space if the field $\phi$ is displaced from its value at the minimum of the potential ($\phi =1$). This can be seen from eq. (\ref{eq:dotH}). This equation implies that if we want to consider the effect of the $R^2$--corrections on preheating, the field $\phi$ has to be displaced from $\phi = 1$ and the expansion rate will be consequently non--zero. Such a constraint does not appear in the Einstein frame, as we will now see. 

The conformal transformation to the Einstein frame is achieved by considering ${\tilde g}_{\mu\nu} = e^{2\psi/\sqrt{6}}g_{\mu\nu}$. Then, choosing 
\begin{eqnarray}
g_{\mu\nu} &=& {\rm diag}\left(-1,a^2(t),a^2(t),a^2(t) \right) \\
{\tilde g}_{\mu\nu} &=& {\rm diag}\left(-1,a^2_E(t_E),a^2_E(t_E),a^2_E(t_E) \right)\;,
\end{eqnarray}
with $dt_E = e^{\psi/\sqrt{6}}dt$, the expansion rate in the Jordan frame $H=\dot a/a$ is related to the expansion rate in the Einstein frame $H_E$, by
\begin{equation}
H_E \equiv \frac{1}{a_E}\frac{d a_E}{d t_E} = e^{-\psi/\sqrt{6}}\left( H + \frac{1}{\sqrt{6}}\dot\psi \right)\;,
\end{equation}
where the dot represents a derivative with respect to $t$. The equation of motion for the fields are 
\begin{eqnarray}
\psi''&+&3H_E \psi'+V_\psi = -\frac{1}{\sqrt{6}} e^{-2\psi/\sqrt{6}}\chi'^2\;, \\
\chi''&+&(3H_E - \frac{2}{\sqrt{6}} \psi')\chi'+e^{2\psi/\sqrt{6}}V_\chi = 0\;,
\end{eqnarray}
where the prime denotes the derivative with respect to $t_E$, $V_\psi = \partial V/\partial \psi$ and $V_\chi = \partial V/\partial \chi$. The Friedmann equation has the standard form in the Einstein frame. It is consistent to neglect the expansion of space (i.e. to set $H_E=0$) and have both fields evolving. In this case the evolution of both fields are still coupled via the kinetic terms. In addtion, the masses of the $\chi$-- and $\sigma$--fields become $\psi$--dependent as well as the coupling $h$, which transforms as $h \rightarrow \tilde{h} = h e^{-\psi/\sqrt{6}}$ and similarly for $m_\chi$ and $m_\sigma$. The evolution of the $\psi$--field, which encodes the modifications of gravity in the Einstein frame, affects the evolution of the $\chi$--field in two ways. Firstly, $\chi$ acts as a source for the oscillations of the $\psi$--field. Secondly, an oscillatory $\psi$--field results in oscillations of the effective masses for $\chi$ and $\sigma$ as well we as the coupling $h$. This is a very different situation from the one studied in \cite{Lachapelle:2008sy}, where the masses and couplings were not functions of $\psi$. In our model, if the amplitude of $\psi$ is not negligible immediately after inflation, the equation for the perturbations of $\sigma$ can no longer be written in Mathieu or Whittaker form and parametric resonance is mitigated.

\section{Conclusion}\label{sec:conclusions}
In this paper we studied in detail the periods of reheating and preheating in a simple extension of the Starobinsky inflationary model, considered previously in relation to the 2015 Planck data \cite{vandeBruck:2015xpa}. We worked with a choice of parameters which lead to predictions consistent with the Planck 2015 data for the primordial power spectra and we focused on non-trivial models, i.e., where the contributions from both the gravitational corrections and the inflaton field are significant at the end of inflation.

In the case of reheating, the mass $m_\phi$ of the scalar degree of freedom associated with the $R^2$--correction appears to have a small effect on the reheating temperature and the duration of reheating. Specifically, when increasing $m_\phi$ (i.e., making the corrections less important), an increase in the reheating temperature and the duration of reheating are observed. While the influence is small for the range of parameter we have considered, it is important to take it into account when comparing the theory to data. The change in the duration of reheating will affect the relation between the $e$--fold number and the wavenumber $k$ of the physical scales, see equation (6) in \cite{Liddle:2003as}. This contribution should not be neglected. 

In the case of preheating, we find that the particle production is much less efficient if the $R^2$--corrections are large at the end of inflation. This is due to the impact these corrections have on the evolution of the expansion rate $H$, which in turn affects the dynamics of the $\chi$ field. We also considered the situation from the perspective of the Einstein Frame, where we point out that parametric resonance is mitigated in our model due to the couplings of the $\psi$--field, which encodes the corrections to Einstein gravity, to the $\chi$ and $\sigma$--fields. Specifically, the masses and couplings of $\chi$ and $\sigma$ become $\psi$--dependent and the oscillating behaviour of $\psi$ influences particle production.

It is interesting to note that modifications of gravity cannot only affect the inflationary epoch itself but also the epoch immediately afterwards. This has been the case for the model studied in this paper and also for the model studied in \cite{vandeBruck:2016xvt}. It is clear from our results that modifications due to gravity have to be taken properly into account also immediately after inflation when comparing the theory to data.

\vspace{0.5cm}

\noindent {\bf Acknowledgements:} CvdB is supported by the Lancaster-Manchester-Sheffield Consortium for Fundamental Physics under STFC grant ST/L000520/1. PKSD is supported by the NRF (South Africa). LEP is supported by a studentship from the School of Mathematics and Statistics at the University of Sheffield. 


\end{document}